\documentclass[letterpaper,aps,prb,twocolumn,amsmath,amssymb,showpacs,
floatfix] {revtex4}

\usepackage{color}
\usepackage[pdftex]{graphicx,hyperref}
\usepackage{dcolumn}
\usepackage{wasysym,marvosym}
\usepackage{times,bm,bbm}

\newcommand{\la}{\langle}
\newcommand{\ra}{\rangle}

\newcommand{\ua}{\uparrow}
\newcommand{\da}{\downarrow}
\newcommand{\bd}{\ensuremath{\bm{\delta}}}
\newcommand{\ce}{\ensuremath{\varepsilon}}	
\newcommand{\bk}{\bm{k}}
\newcommand{\ACdir}{\ensuremath{\cal AC}}
\newcommand{\ZZdir}{\ensuremath{\cal ZZ}}
\newcommand{\tav}{{\langle t \rangle}}
\def\lapp{\lower.35em\hbox{$\stackrel{\textstyle<}{\sim}$}}

\newlength{\textwidthm}

\begin{document}

\title{Magnetism in strained graphene dots}

\author{J. Viana-Gomes$^1$, Vitor M. Pereira$^2$, and N. M. R. Peres$^1$} 

\affiliation{$^1$ Department of Physics and Center of Physics,
 University of  Minho, P-4710-057, Braga, Portugal}

\affiliation{$^2$ Department of Physics, Boston University, 
590 Commonwealth Avenue, Boston, Massachusetts 02215, USA}

\date{\today}

\begin{abstract}
We study the magnetization of square and hexagonal graphene 
dots. It is shown that two classes of hexagonal dots have a second
order phase transition at a critical Hubbard energy $U$, whose value
is 
similar to the one in bulk graphene, 
albeit the dots do not have a density of states
proportional to the absolute value of the energy, relatively to the
Dirac point. 
Furthermore, we show that a particular class of hexagonal dots
having zig-zag edges, does not exhibit zero energy edge states.
We also study the effect of uniaxial strain on the evolution of the
magnetization of square dots, and find that the overall effect is an
enhancement of magnetization with strain. The enhancement can be as
large as 100\% for strain of the order of 20\%. Additionally, stress
induces a spatial displacement of the magnetization over the dot,
moving it from the zig-zag to the armchair edges.

\end{abstract}


\maketitle

\section{Introduction}

Nowadays, the terms {\it wonder material} and {\it graphene dreams}
\cite{ScienceDreams,EPN} frequently accompany the description
of the unusual electronic \cite{Nov04,pnas,rmp}, thermal \cite{lau},
and mechanical properties of graphene \cite{Ni,Hone,HeeHong}. 
One of the most promising {\it graphene dreams} is its application 
to a new generation of nanoelectronic devices \cite{westervelt}. 
To that effect, a number of systems have already been experimentally
investigated, namely: 
single-electron transistors \cite{Bunch},
quantum interference devices \cite{Miao},
and graphene dots \cite{Ensslin,Ponomarenko}.
The presence of Coulomb oscillations in graphene quantum dots was also 
identified by different groups \cite{Bunch,Ensslin,Ponomarenko}.

Theoretically, the first investigations in this context focused on the
transport properties of short (and wide) ribbons \cite{Beenakker06}. For
long  graphene ribbons \cite{Palacios}, it was shown that the low
bias current flowing through the bulk of the ribbon is very robust
with respect to a variety of constriction geometries and edge defects,
a result also confirmed for disordered armchair nanoribbons
\cite{Wakabayashi2009}. 
As in the case of short ribbons \cite{Beenakker06}, evanescent waves
were seen to play an important role in the electronic transport
through graphene quantum dots \cite{Guinea}.
The role of magnetic fields in the electronic levels of graphene
quantum dots has been investigated by several authors. Of particular
interest for transport properties is the fact that optical properties
can be tuned by the size and edge type of the dot \cite{Peeters}.
The shape and symmetry \cite{Baranger,Manninen} of the dots also play
an important role on energy level statistics and charge density.
For the special case of triangular quantum dots \cite{Manninen},
the existence of ``ghost states'' was revealed, when these dots have
armchair edges, whereas for triangular dots with zigzag edges the well
known surface states are present.
Of particular importance was the demonstration of large insensitivity 
of the electronic structure to the edge roughness \cite{Manninen}.

The main motivation for research in graphene quantum dots and ribbons
is related to the need of producing a graphene-based system with an
energy gap, which is not present in bulk graphene. This fact is a
recognized shortcoming of bulk graphene, in what concerns applications
relying on current electronic operation.
Gaps can be induced by electronic quantum confinement in narrow
armchair ribbons \cite{Wakabayashi}, a result confirmed by {\it
ab-initio} calculations \cite{Louie}, and experimentally
\cite{Kim-Ribbons}. 
First principle calculations further show that zig-zag ribbons can
support magnetic ground states which leads to a gapped 
spectrum \cite{Louie}. Spin polarized ground states are also present
in small graphene derivative molecular systems \cite{Veronica}. This
finding opens a new line of research: the study of spin polarized
ground states of graphene quantum dots of different geometries. In
both single \cite{Louie,Veronica} and bilayer \cite{evcastromag}
zig-zag ribbons, it was found that opposite edges align
antiferromagnetically, with the magnetization rapidly decaying towards
the bulk of the ribbon.  Hartree-Fock (HF) calculations are specially
suited for this type of study, since one can study the effect of
different values of the Coulomb interaction on the magnetic structure.
An interesting effect emerges from such studies \cite{evcastromag}: 
graphene ribbons have no critical Hubbard interaction, $U$, and thus
the HF ground state is always magnetic. This result is at odds with
the behavior of bulk graphene \cite{Sorella,Martelo,Bozi,Paiva}
and bilayer graphene \cite{evcastromag,CastroPRL,CastroJPCM}.
This richness of different behaviors suggests studying the formation of
magnetic ground states in graphene dots, a line a research we carry on
in this paper from an HF point of view.

Interest in magnetism of sp$^2$ carbon systems was greatly spurred by
experiments with proton-irradiated graphite \cite{Esquinazi_a}, and
with the experimental evidence that the measured magnetism might stem
from $\pi$-orbital physics alone \cite{Esquinazi_b}. Proton
irradiation induces spatially disordered vacancies in the system
\cite{peres,PereiraVacancy}. The magnetism found experimentally is
supported theoretically by Hartree-Fock and \emph{ab-initio} studies
\cite{Oleg}. 
Recent experimental developments addressed the intrinsic
ferromagnetism in HOPG graphite, originating from the naturally
occurring grain boundaries, where zig-zag edges develop and local
magnetic moments are formed. Typical hysteretical curves of a
ferromagnetic material are seen in a temperature range from 5~K up to
300~K \cite{Magnetism-naturephys}.

Hence, disorder, such as those line defects studied recently, is 
a possible route for ferromagnetism in carbon-based materials.
However, disorder is not a necessary condition for magnetism in sp$^2$
systems. It is by now well established theoretically that graphene
systems with zig-zag edges can support magnetic moments and, in a
system with perfect edges, this leads to magnetic ground states. An
argument widely used against this result is based on the fact that the
atoms at the edges are, essentially, gigantic free radicals, which
would be impossible to realize in a true graphene system. This
argument ignores, however, the fact that such {\it gigantic free
radicals} can be chemically passivated with other chemical species,
notably hydrogen. It is found \emph{ab-initio} that the long range
magnetic order is robust, even under passivation of the edges
\cite{Louie}, confirming early predictions \cite{fujita96}.

In small graphene structures (triangles and hexagons),  
magnetism has been thoroughly investigated \cite{palacios_a} both
using {\it ab-initio} and Hartree-Fock methods, but generally for
small dot sizes. The interplay between transport and magnetism has
also been addressed \cite{palacios_b}, as well as magnetism induced by
vacancies \cite{palacios_c} (which can be seen as a three site zig-zag
edge). 
 
Another topic of experimental research that has recently seen a
considerable upsurge, is the study of interplay between the mechanical
properties of graphene and its electronic structure. The motivation
for these studies is the possibility of tailoring the transport
properties of graphene by means of externally induced strain
\cite{OrigamiPereira}.
Naturally related is the question of how can the above mentioned
magnetic properties of graphene ribbons and dots be modified by
external stress. 
In a previous work \cite{StressPereira}, some of the present
authors showed that the electronic spectrum of graphene can be strongly
modified by external stress. In particular, stress along the zig-zag
edges of the system might eventually lead to the opening of a gap at
large deformations. In addition to studying the magnetic properties of
quantum dots in equilibrium, here we will also address how their
magnetization is affected by external stress.

Our main findings can be summarized as follows. The existence of
magnetic ground states in graphene dots of nano to mesoscopic sizes
depends on their geometry, and not only on the existence of zig-zag
portions along their edges. Within a Hartree-Fock framework, the
existence (or not) of a minimum on-site Coulomb repulsion, $U_c$, for
the onset of magnetism depends critically on the dot geometry and
symmetry. When strain is applied, the nearest-neighbor hopping
integrals are naturally modified. This leads to a modification of the
local magnetic moments found in the ground state, in a way which is
much stronger than one would expect just by calculating
the isotropic renormalization of the critical Coulomb repulsion
$U$. Our results show that magnetism is enhanced under
uniaxial strain, and causes a reduction of $U_c$, for the
dots which exibith finite $U_c$.
Moreover, we find that, under strain, the local magnetic moments
associated with zig-zag edges in rectangular dots can drift from the
zig-zag to the armchair edges.

The paper is organized as follows: in Sec. \ref{model} we introduce
our theoretical model, and discuss the relevance of several Coulomb
terms in defining an effective Coulomb interaction $U$. A discussion
of the appropriate value of $U$ for graphene ensues. In Sec.
\ref{sec:magnetization} we study the magnetization of different types
of square and hexagonal graphene dots. In Sec. \ref{sec:strain} the
role of strain on the magnetization of graphene dots is considered.
Our main results are discussed in Sec. \ref{sec:disc}.

\section{Model}
\label{model}

Our study of magnetism in graphene quantum dots and antidots relies on
the Hubbard model with on-site interaction, an approach used
by other authors in the study of graphene ribbons \cite{Oleg}. 	 
Since dots have no translation symmetry, the problem is
solved in real space. To that end, we need to set up the Hamiltonian
in a matrix form, which requires a convenient algorithm to build such
a matrix for the different type of dots.
In what follows we describe the model, together with the 
physically relevant values of the on-site Coulomb interaction.

\begin{figure}[tb]
  \centering
  \includegraphics*[width=0.35\textwidth]{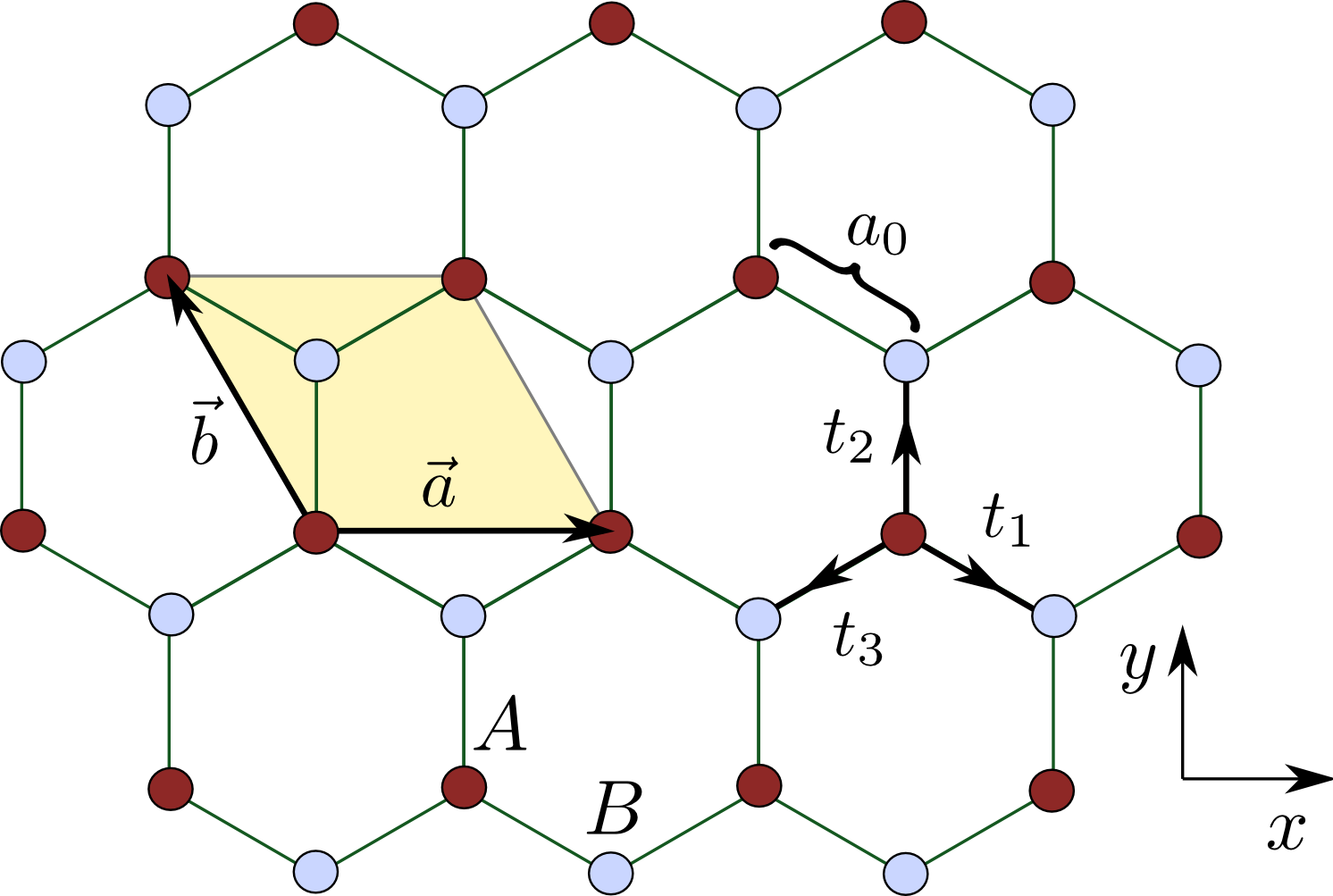}
  \caption{(Color online) 
    Illustration of the honeycomb lattice with the $A$ and
    $B$ sublattices, the lattice vectors $\bm\delta_i $ ($i=1,2,3$),  
    the primitive vectors $\bm a$ and $\bm b$, and the
    hoppings $t_i$ ($i=1,2,3$) used in Sec. \ref{sec:strain}. 
    The abscissas are along the zig-zag edge (horizontally in the figure).
    and $a_0$ is the equilibrium carbon-carbon distance.}
\label{fig:lattice}
\end{figure}

The study of magnetism in condensed matter physics is traditionally, and
frequently, based upon the Hubbard model, which can be written as
\begin{eqnarray}
H&=&H_0+H_U\,,\\
H_0&=& -t\sum_{\bm r,\bm\delta,\sigma=\uparrow,\downarrow}
a^\dag_\sigma(\bm r)b_\sigma(\bm r+\bm\delta)+{\rm H.c.}\,,\\
H_U&=&U\sum_{\bm r}a^\dag_\ua(\bm r)a_\ua(\bm r)
a^\dag_\da(\bm r)a_\da(\bm r)\,,\nonumber\\
&+&U\sum_{\bm r}b^\dag_\ua(\bm r)b_\ua(\bm r)
b^\dag_\da(\bm r)b_\da(\bm r)\,.
\end{eqnarray}
For graphene the hopping integral is $t\simeq 2.7$ eV (used as the
energy unit in this work), $U$ is the on-site Coulomb repulsion
energy, and $a^\dag_\sigma(\bm r)$ [$b^\dag_\sigma(\bm r)$] is the
electronic creation operator at site $A$ [site $B$] of the unit cell
$\bm r$ in the honeycomb lattice (see Fig. \ref{fig:lattice}).

The Coulomb term is treated at the mean-field level
\cite{Bozi,evcastromag} by making the replacement of the quartic
interaction by
\begin{eqnarray}
 H_U\rightarrow H_U^{MF}&=&
U\sum_{\bm r,\sigma}a^\dag_\sigma(\bm r)a_\sigma(\bm r)
\la a^\dag_{-\sigma}(\bm r)a_{-\sigma}(\bm r)\ra\,,\nonumber\\
&+&U\sum_{\bm r,\sigma}b^\dag_\sigma(\bm r)b_\sigma(\bm r)
\la b^\dag_{-\sigma}(\bm r)b_{-\sigma}(\bm r)\ra\,,
\label{HUMF}
\end{eqnarray}
such that, when $\sigma=\ua,\da$,  we have $-\sigma=\da,\ua$.
After this transformation, the quantum problem becomes
bi-linear in the electronic operators, and can be solved by 
diagonalization of two matrices of dimension $D\times D$, where $D$ is
the total number of lattice sites in the dot. The electronic
density has to be determined self-consistently and the mean-field
equations read \cite{Hirsh}
\begin{eqnarray}
\label{MFeqa}
 n_{a,\sigma}(\bm r)&=& \la a^\dag_{\sigma}(\bm r)a_{\sigma}(\bm r)\ra\,,\\
\label{MFeqb}
 n_{b,\sigma}(\bm r)&=& \la b^\dag_{\sigma}(\bm r)b_{\sigma}(\bm r)\ra\,.
\end{eqnarray}
Here $n_{a,\sigma}(\bm r)$ and $n_{b,\sigma}(\bm r)$ are the mean
electronic densities of spin $\sigma$ at the $A$ and $B$ sites of the
unit cell $\bm r$, respectively. The  wave function of the system
corresponding to an energy $E_{\lambda,\sigma}$ is labeled by the
quantum number $\lambda$, having the explicit form
\begin{equation}
 \vert\psi_{\lambda,\sigma}\ra=\sum_{\bm r}A_{\lambda,\sigma}(\bm r)
\vert a,\bm r\ra+ B_{\lambda,\sigma}(\bm r)
\vert b,\bm r\ra\,,
\label{WF}
\end{equation}
where $\vert a,\bm r\ra$ and $\vert b,\bm r\ra$ are lattice-position
basis-states. The mean field equations (\ref{MFeqa}) and (\ref{MFeqb})
are determined as function of the $A_{\lambda,\sigma}(\bm r)$ and
$B_{\lambda,\sigma}(\bm r)$ coefficients, defined in the wave function 
(\ref{WF}), according to
\begin{eqnarray}
 n_{a,\sigma}(\bm r) =\sum_{\lambda}\vert A_{\lambda,\sigma}(\bm r)\vert^2
f(E_{\lambda,\sigma})\,,\\
  n_{b,\sigma}(\bm r) =\sum_{\lambda}\vert B_{\lambda,\sigma}(\bm r)\vert^2
f(E_{\lambda,\sigma})\,,
\end{eqnarray}
where $f(x)=(1+e^{\beta(x-\mu)})^{-1}$, $\mu$ is the chemical
potential and $\beta=1/(k_BT)$, with $T$ the temperature. The problem
has to be solved numerically. We start with a trial solution for
$n_{a,\sigma}(\bm r) $ and $n_{b,\sigma}(\bm r) $; then the
Hamiltonian is diagonalized and new values for $n_{a,\sigma}(\bm r) $
and $n_{b,\sigma}(\bm r) $ are computed; the procedure is iterated a
number of times until convergence is reached.

As mentioned, the conventional treatment of magnetism in graphite and
graphene at the Hartree-Fock level includes only the effect of the
on-site Coulomb interaction $U$. We now discuss the importance of more
general interactions \cite{Vollhardt}. We first note that, at the
mean-field level, a nearest neighbor Coulomb interaction does not
contribute to the existence of a ferromagnetic phase in the case of a
system with translational invariance and a single orbital per unit
cell \cite{kampf}. If in graphene we consider a Coulomb term of the
form
\begin{equation}
 H_V=
V\sum_{\bm r,\bm\delta,\sigma,\sigma'}a^\dag_\sigma(\bm r)a_\sigma(\bm r)
b^\dag_{\sigma '}(\bm r+\bm\delta)b_{\sigma'}(\bm r+\bm\delta)\,,
\end{equation}
it remains true that such interaction will not contribute (at the
Hartree-Fock level) to the existence of a magnetic ground state in the
thermodynamic limit. The situation is different, though, in a system
without translational invariance, since the spin density in
neighboring carbon atoms is not necessarily equal. This is of special
relevance near the edges of the system.

The mean field Hamiltonian has the form
\begin{equation}
H_V\rightarrow H_V^{MF}=V\sum_{\bm r,\sigma} a^\dag_\sigma(\bm r)a_\sigma(\bm r)
\bar n_b(\bm r)+(a\leftrightarrow b)\,,
\label{HVmf}
\end{equation}
where $(a\leftrightarrow b)$ in Eq. (\ref{HVmf}) is a shorthand notation
for a term with the same form as the first, but with the role of the
$a$ and $b$ operators interchanged, and 
\begin{equation}
\bar n_b(\bm r)= \sum_{\bm\delta,\sigma'}
\la b^\dag_{\sigma'}(\bm r+\bm\delta)b_{\sigma'}(\bm r+\bm\delta)\ra\,,
\end{equation}
is the average density at the $B$ neighbor carbon atoms of a given $A$
atom at position $\bm r$. The terms $H_U$ and $H_V$ are direct Coulomb
interactions. An exchange term can also be included in the
Hamiltonian, having the form
\begin{equation}
 H_J=\frac J 2\sum_{\bm r,\bm\delta}\bm S_a(\bm r)\cdot\bm S_b(\bm r+\bm\delta)\,,
\end{equation}
with $S_a(\bm r)$ [$S_b(\bm r)$] the electronic spin operator of an
electron at site $\bm r$ of the sub-lattice $A$ [$B$]. In this case,
the mean field Hamiltonian is
\begin{equation}
H_J\rightarrow H_J^{MF}=\frac J 2
\sum_{\bm r,\sigma}  a^\dag_\sigma(\bm r)a_\sigma(\bm r)
\bar\Sigma_b(\bm r)+(a\leftrightarrow b)\,,
\label{HJmf}
\end{equation}
with
\begin{equation}
\bar \Sigma_b(\bm r)=\sum_{\bm\delta,\sigma'}
\sigma'\la b^\dag_{\sigma'}(\bm r+\bm\delta)b_{\sigma'}(\bm r+\bm\delta)\ra\,,
\end{equation}
where $\Sigma_b(\bm r)$ is the average spin density at the $B$
neighbor of a given $A$ atom at position $\bm r$, and $\sigma$ takes
the values $\pm 1$ when used as a multiplicative factor. 

We shall assume that the leading overall effect of these three
interactions can be captured by a renormalized Hubbard interaction,
$U$, in the mean field calculations. Thus the value of $U$ should
reflect this effective interaction, rather than the bare on-site
Coulomb repulsion in graphene.

We now proceed to study the ground state of dots and their
magnetization as a function of $U$. A natural question immediately
arises: what value of effective $U$ should one take to be consistent
with the magnitude of the real Coulomb interactions in the material.
For the benzene molecule $U$ was seen to be as large as 16~eV
\cite{Parr}. In a recent study of magnetism in disordered graphene and
irradiated graphite \cite{Oleg}, the value of $U$ was considered to be
in the interval 3-3.5 eV, based on the value accepted for {\it
trans-}polyacetylene, a one-dimensional bipartite sp$^2$ carbon system
(although this value of $U$ for {\it trans-}polyacetylene has been
subject to controversy \cite{Baeriswyl}). Other two recent studies
\cite{palacios_a,Jung} took $U=$2 eV and $U=3.85$ eV,  values that reproduce 
the LDA gap in
graphene ribbons \cite{Louie} and the
HOMO-LUMO gap in small graphene based structures. 
We shall consider below  $U=2$ eV and $U=3.5$ eV
as reference values in our calculations.

\section{$\Delta(N)$ and magnetization}
\label{sec:magnetization}

The results for the magnetization of graphene dots 
depend on the type of edges present. Generally speaking, one expects
to see larger magnetization close to zig-zag edges, where the
existence of localized states satisfies a spatial Stoner
criterion for finite values of $U$ \cite{Vozmediano}. The
existence of such type of states is shown to be related to lattices
with an odd number of sites \cite{mudry}. For models with
sub-lattice symmetry, as is the case of graphene, the number
of zero energy modes is determined from the difference
$\vert N_A-N_B\vert$, were $N_A$ and $N_B$ is the number
of sites in sub-lattice $A$ and $B$ respectively
\cite{mudry,Pereira-Disorder}. 

In our calculations we use relatively small dots. This choice is
justified because there are almost no visible finite-size effects,
as we explicitly show below by studying dots of different sizes.
Additionally, our choice is also justified from an experimental point
of view, since it recently became possible to cleave graphene
crystalites down to one-dimensional chains by irradiation inside a
transmission electron microscope [see Fig. 2 of Ref.
\onlinecite{1DchainsIijima}]. 
With such new experimental methods, tailoring dots of any possible
size and shape seems now quite within reach.

It is useful for latter use to introduce the quantity $\Delta(N)$, as
the energy interval between the highest hole state and the first
particle one, for the system without interactions ($U=0$):
\begin{equation}
 \Delta(N)=E^{\rm particle}_{\rm lowest}-E^{\rm hole}_{\rm highest}\,,
\end{equation}
where $N$ is the total number of atoms in the dot. We consider two
types of dots, with square and hexagonal shapes, and also the case of
a dot with two non-connected regions (some times referred to as an
anti-dot). 

\begin{figure}[tb]
  \centering
  \includegraphics*[width=0.4\textwidth]{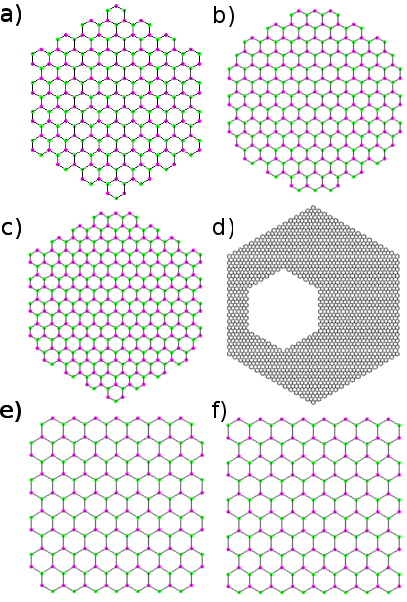}
  \caption{(Color online) 
    Types of hexagonal and square dots studied in this
    work: (a) hexagon with $D_6$ symmetry and no zig-zag
    edges (termed HEX2 in the figures below);
    (b) hexagon with   $D_6$ symmetry and zig-zag
    edges (termed HEX1 in the figures below);
    (c) hexagon with $D_3$ symmetry with zig-zag
    edges; (d) an anti-dot with the external and internal
    boundaries made of $D_6$ symmetry with no zig-zag edges;
    (e) square where the vertices are of zig-zag type  (termed SQR1);
    (f) square where the vertices are of armchair type (termed SQR2).
    The size of the figures is characterized by the number $L$
    of carbon atom horizontal-lines (zig-zag type of lines); 
    for example, in panel (a) one has $L=14$, and in panels (e) and
    (f)  one has in both cases
    $L=10$.
  }
  \label{fig:dots}
\end{figure}

\begin{figure}[tb]
  \includegraphics*[width=0.45\textwidth]{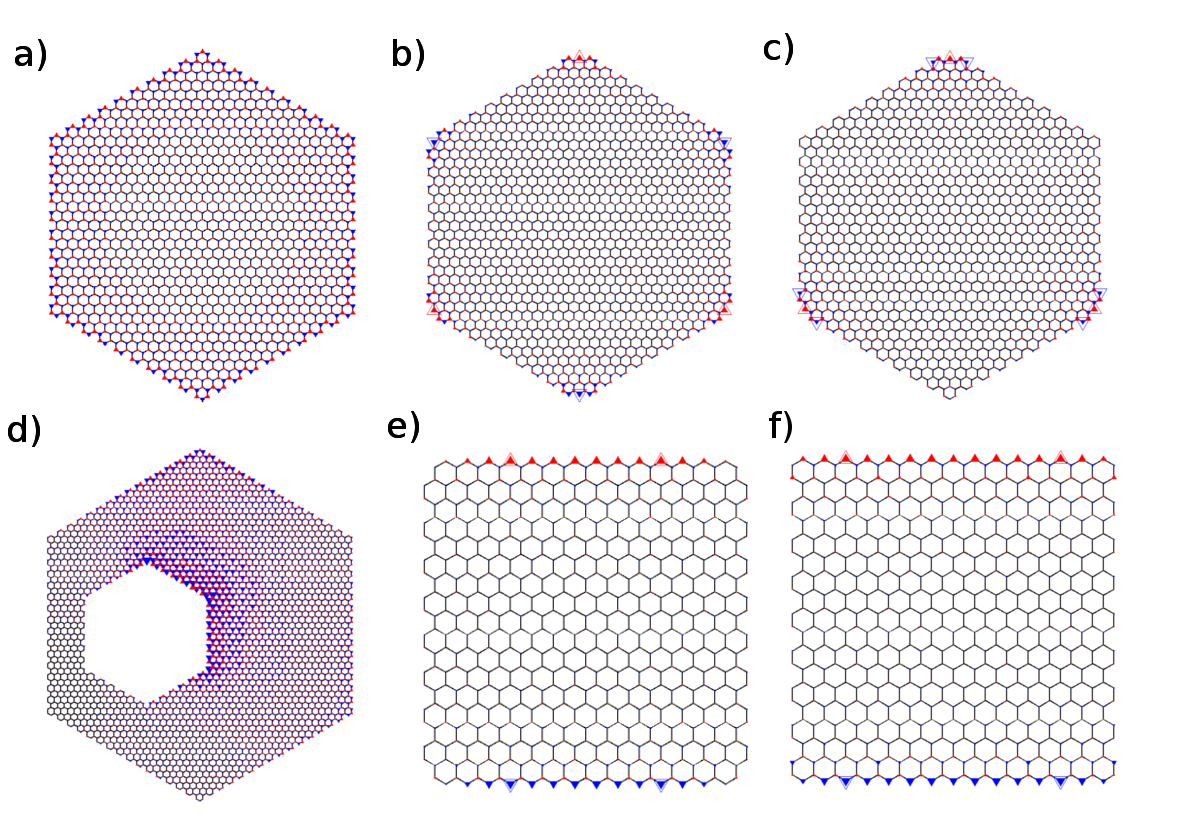}
  \caption{(Color online) 
    Illustration of the local magnetization for dots of the
    same type illustrated in Fig. \ref{fig:dots}, but with a much
    larger number of atoms, thus avoiding finite-size effects.
    Upright triangles refer to positive magnetization and the
    down ones to negative values. The open triangles refer to the
    points where the magnetization has it maximum value.
    In panel (a) we have a HEX2 hexagon, with $U=2.3$;
    for (b) and (c) hexagons $U=2$. In panel (d) we show
    an anti-dot, where the external boundary is
    from a EHX2 hexagon and the internal one is from a HEX1
    hexagon. Panels (e) and (f) are of type SQR1 and SQR2
    respectively.
  }
\label{fig:hex}
\end{figure}

\begin{figure}[tb]
  \centering
  \includegraphics*[width=0.45\textwidth]{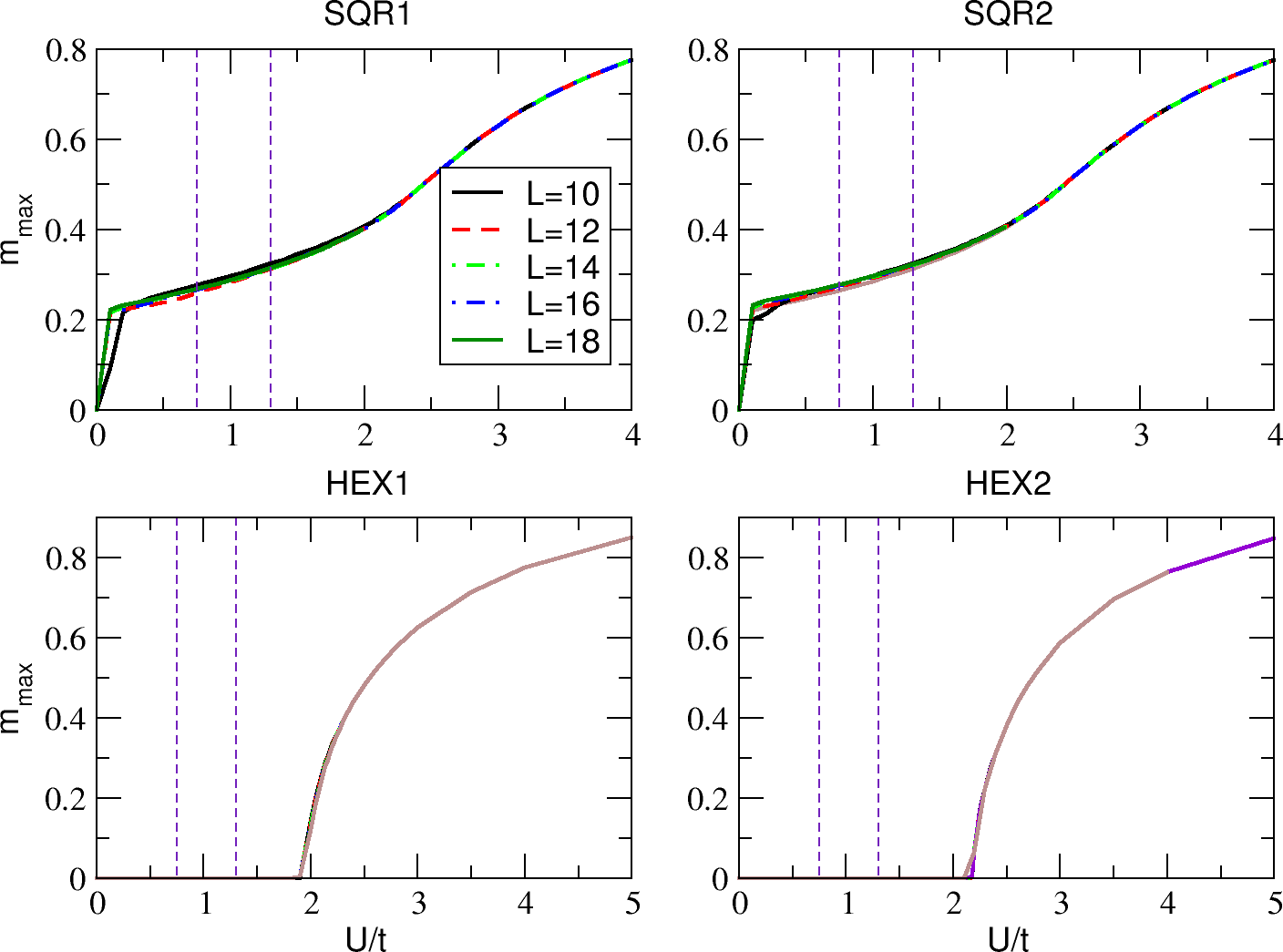}
  \caption{(Color online) 
    Variation of the magnetization as function of $U$, for
    squares and hexagons of different types. The vertical dashed lines
    refer to the values of $U$ used in Refs. \onlinecite{Oleg,Jung}
    (see text in Sec. \ref{model} for a discussion about these
    choices).
  }
  \label{fig:magU}
\end{figure}

We start with the study of hexagonal dots. There are hexagonal dots with
different symmetries and different types of edges:
\begin{enumerate}
  \item dots with $D_6$ symmetry, having only armchair edges
    (see Fig. \ref{fig:dots} a) ).
  \item dots with $D_6$ symmetry, having armchair and zig-zag edges
    (see Fig. \ref{fig:hex} b) ).
  \item dots with $D_3$ symmetry, having armchair and zig-zag edges
    (see Fig.\ref{fig:dots} c) ).
\end{enumerate}
The first type of dot defined above shows that it is possible to have
dots without zig-zag edges, no matter how large they are, and
therefore the physics associated with zig-zag edges should not be
present. This type of $D_6$ dot, when very large, is almost equivalent
to the bulk system, having the full symmetry of the honeycomb lattice
and therefore showing a second order phase phase transition at a
(mean-field) critical Hubbard interaction, $U_c$, given by
\begin{equation}
\label{Uc}
U_c\simeq 2.23 t\,,
\end{equation}
as shown in Fig. \ref{fig:magU} (HEX2-type). The same holds true for
the $D_6$ HEX1-type of hexagons, but with a smaller value of $U_c$
(smaller than 2). The dependence of the maximum value of magnetization
as a function of $U$ for the two $D_6$ hexagons is plotted in Fig.
\ref{fig:magU}. There we see that the critical $U$ is close to that
given by Eq. (\ref{Uc}), without any noticeable variation with the
size, $L$, of the hexagon. In Fig. \ref{fig:magU} the reference values
for $U$ discussed at the end of Sec. \ref{model} are represented as
vertical dashed lines. Clearly, the magnetic transition is well above
those reference values for $U$, meaning that this type of dots, if
experimentally fabricated, should exhibit no magnetic order.

\begin{figure}[tb]
  \includegraphics*[width=0.45\textwidth]{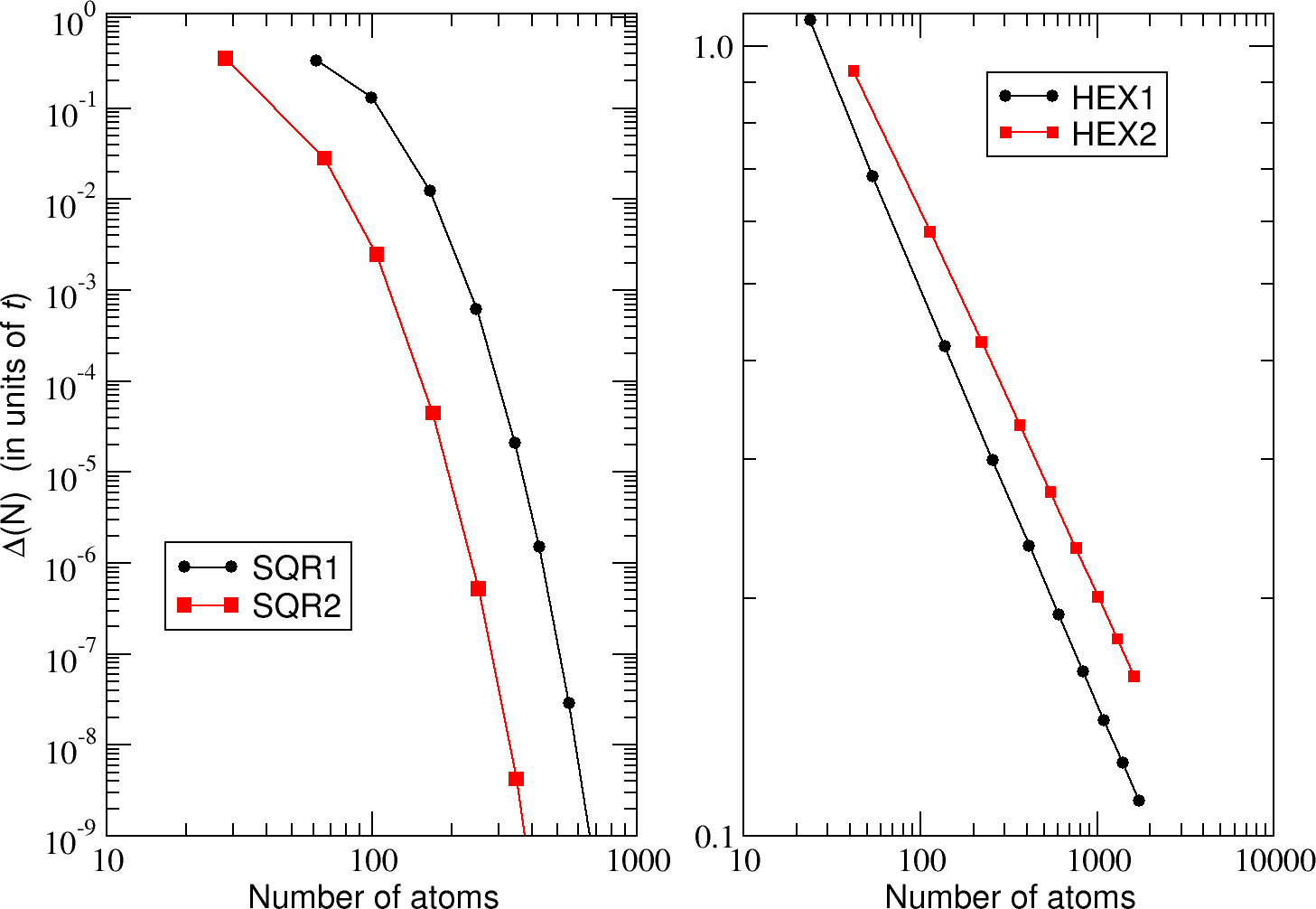}
  \caption{(Color online) 
    Variation of quantity $\Delta(N)$ with the number of atoms 
    for square and hexagonal dots. From this figure alone we can
    understand that the spectrum of hexagons of type HEX1 and HEX2
    behaves exactly in the same way, not showing the development of
    zero energy edge states, a consequence of the D6 symmetry alone.
  }
  \label{fig:gap}
\end{figure}

We note that the two $D_6$ hexagons have finite $\Delta(N)$ values,
which vary as a power law with $N$, as shown in Fig. \ref{fig:gap}.
For the HEX1- and EXH2-types of hexagons we numerically extract:
\begin{eqnarray}
  &&\Delta(N)\simeq 1.71N^{-0.53}\,,\\
  &&\Delta(N)\simeq 1.75N^{-0.48}\,,
\end{eqnarray}
respectively. The exponent in the above power laws is essentially
equal to $\frac{1}{2}$, and, therefore, reflects the finite-size
quantization of the electronic spectrum. For square dots, on the other
hand, we find that $\Delta(N)$ vanishes much rapidly as $N$ increases,
reflecting the formation of edge states at nearly zero energy: for
small systems, the edge states from opposite sides of the square dot
hybridize, and the otherwise zero energy states for the semi-infinite
system split in energy. As the width of the dot increases the
hybridization is strongly suppressed and zero energy levels develop.
 
The finiteness of $\Delta(N)$ for the hexagons correlates 
with the finite value of $U_c$ seen in Fig. \ref{fig:magU}.
On the other hand, the value $U_c\simeq 2.23 t$ previously obtained in
the literature \cite{Sorella} was determined using the fact that the
density of states of bulk graphene is proportional to the absolute
value of the energy relatively to the Dirac point, being zero for a
half-filled system. These two results -- for $D_6$ hexagonal clusters
and the bulk system -- means that the value of $U_c$ (\ref{Uc}) is not
exclusively determined by the vanishing nature of the density of
states at the Dirac point of bulk graphene.
On the other hand, the two $D_6$ hexagons show different values of
$U_c$, which can only be interpreted as a boundary effect, determined
by the different nature of their edges. It is worth noticing that the
hexagons of type HEX1, having zig-zag terminations (defining a figure
with $D_6$ symmetry) do not develop zero energy states, leading,
therefore, to the finiteness of $U_c$. The hexagonal dot of $D_3$
symmetry shows a behavior for $\Delta(N)$ identical to that found for
the squares and, as a consequence, there is no finite value of $U_c$:
the system is magnetic for any arbitrarily small value of $U$.

The mean-field values of $U_c$ determined for the $D_6$ hexagons will
be modified by quantum fluctuations. The  effect of quantum
fluctuations amounts in general to shifting the Hartree-Fock $U_c$ to
higher values \cite{Sorella,Martelo,Paiva}. In the case of small
graphene-based nanodots, such as bisanthrene \cite{Veronica}
(C$_{28}$H$_{14}$), the Hubbard Coulomb interaction may be larger than
that assumed for macroscopic sp$^2$ carbon systems. This hypothesis is
based on the value $U\sim 16$ eV computed for benzene \cite{Parr}.
Given this value and our current results, there is a real possibility
of having magnetic ground states in small hexagonal systems with $D_6$
symmetry.

In what concerns the relation between magnetism and edge structure, we
see that dots with $D_3$ and square symmetry have zig-zag edges, and
this leads to finite magnetization for any finite $U$. Magnetization
is maximal at, and close to, the zig-zag edges, and fades rapidly as
one progresses towards the bulk of the dot. It is worth noticing that,
for HEX1-type hexagonal dots, there are six external zig-zag
boundaries, but its spectrum does not present zero-energy eigen
values.
In Fig. \ref{fig:hex} we present particular cases of the spatial
distribution of magnetization in the different dots considered in Fig.
\ref{fig:dots}. For the hexagons of type HEX2 the magnetization is
homogeneous over the boundary and, as soon as $U>U_c$, it develops
from the boundary of the hexagon toward the bulk. For HEX1 hexagons
the maximum of magnetization develops at the zig-zag vertices, but
again only for $U>U_c$. In the case of the hexagon with $D_3$ symmetry
the development of the magnetization follows the pattern of that found
for HEX1 dots, but it is finite for any finite $U$ value. The antidot
case (panel (d) of Fig. \ref{fig:hex}) can be considered a simple case
of a disordered system, since all symmetries are broken. In this case
the magnetization develops preferentially at the internal edges
connected to the bulk of the system, with formation of a {\it shadow}
region at bottom left of the anti dot where no magnetization is seen
(for that particular value of $U$). This behavior can be understood
since the internal boundary plays, in the anti-dot, the same role as
the external boundary of the equivalent dot. It is worth noticing that
the anti-dot, being made of HEX1 and HEX2 hexagons has a critical
Hubbard interaction, which is controlled by the $U_c$ value of the
HEX1 hexagon.

In Fig. \ref{fig:magU} we depict the dependence of the maximum of
magnetization ($m_{\rm max}$) with $U$. We see that for the square
dots of both types considered in Fig. \ref{fig:dots} the magnetization
is finite down to arbitrarily small values of $U$. This can be
correlated with the correspondingly small values of $\Delta(N)$, shown
if Fig. \ref{fig:gap}, and the behavior of the DOS at the Fermi level.
As to the hexagonal dots (see Fig. \ref{fig:magU}), we also see that
the existence of a finite $\Delta(N)$ is associated with the existence
of a finite $U_c$. In other words, some geometries have finite density
of states at $E=0$, $\rho(0)\ne0$, whereas others do not. In the case
$\rho(0)=0$ and as $N\rightarrow\infty$ the behavior of the system is
essentially that of the bulk case up to finite-size corrections. On
the other case, with $\rho(0)\ne0$ for finite $N$, the magnetic
behavior is different and there is finite magnetization for any value
of $U$.
We then understand the result obtained by Sorella and Tosatti
(\cite{Sorella}) as the limiting case of $N\rightarrow\infty$ with
$\rho(0)=0$ for any finite $N$.  

\begin{figure}[tb]
  \centering
  \includegraphics*[width=0.45\textwidth]{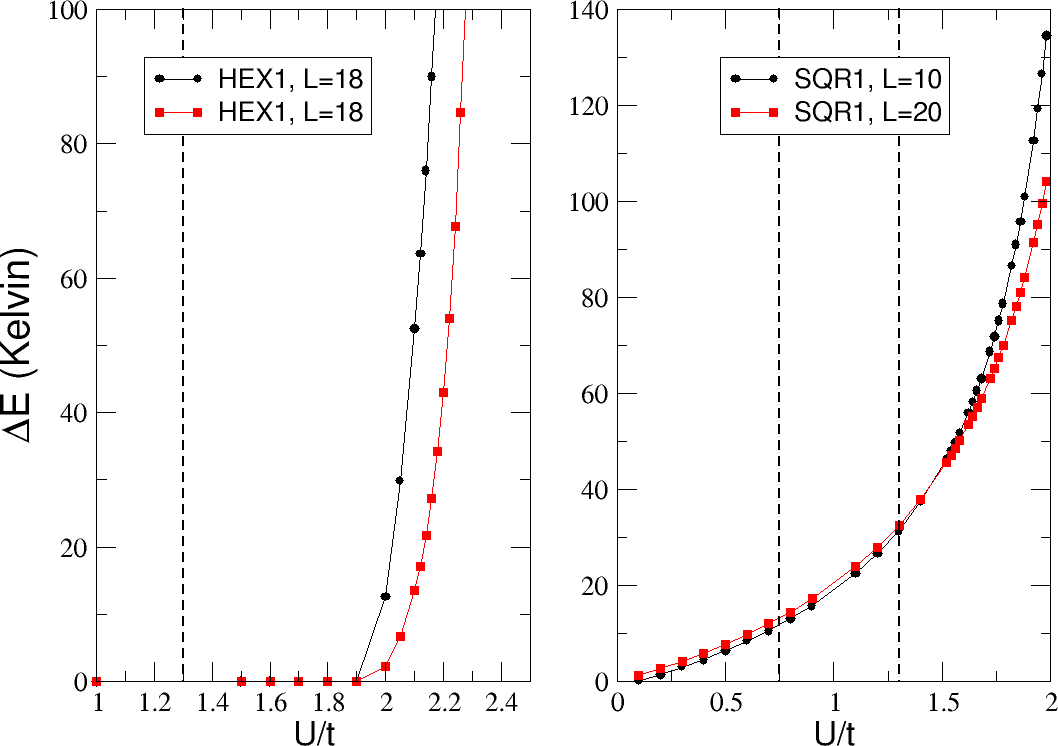}
  \caption{(Color online) 
    Variation of $\Delta E$ with the Coulomb interaction $U/t$,
    for hexagons of the type HEX1, and squares of the type SQR1.
  }
  \label{fig:energy}
\end{figure}

A relevant quantity to compute is the energy difference between the
paramagnetic and ferromagnetic ground states, defined as
\begin{equation}
  \Delta E = E_{\rm para} - E_{\rm ferro}\,, 
\end{equation}
where $E_{\rm para}$ and $E_{\rm ferro}$ are the ground state
energies of the paramagnetic and ferromagnetic ground states. The
value of $\Delta E$ is intimately related to the (mean-field)
temperature at which such a magnetic ground state could be observed.
In Fig. \ref{fig:energy} the value of $\Delta E$ is given in Kelvin
for both hexagons and squares. As discussed previously, these squares
magnetize for any finite $U$ and therefore we could, in principle,
observe magnetism with the values of $U$ expected for graphene (dashed
lines in Fig. \ref{fig:energy}). The mean-field critical temperature
is relatively high, between 10 K and 30 K, in the interval for the
lower and higher expected $U$ in graphene, but not comparable to the
room temperature magnetism observed in graphite
\cite{Magnetism-naturephys}.

\section{Magnetism and strain}
\label{sec:strain}

Strain in graphene in now an active topic of experimental
research. It was shown that some amount of strain 
can be induced either by deposition of oxide capping layers \cite{Ni}
or by mechanical methods \cite{Lombardo}. The amount of strain
can be determined by monitoring the blue \cite{Ni} or
red \cite{Lombardo} shifts of the $G$ and $2D$ Raman peaks of graphene.
This method is a straightforward extension of related studies used in
graphite nano-fibers\cite{sakata}.
Strain has also obvious consequences on the electronic and heat
transport, producing metal-semiconductor transitions, as in carbon nanotubes
\cite{Minot}, or transport anisotropy in graphene \cite{SooKim}. These effects
are due to changes in the band-structure of the materials as a consequence
of the modification of inter-atomic distances, which in turn implies
a change of the electronic hopping parameters. To our best knowledge, the
first correct studies of strain effects on the bandstructure of graphene
were undertaken in Ref.~\onlinecite{StressPereira,RicardoStrain}.

\begin{figure}[tb]
  \centering
  \includegraphics*[width=0.45\textwidth]{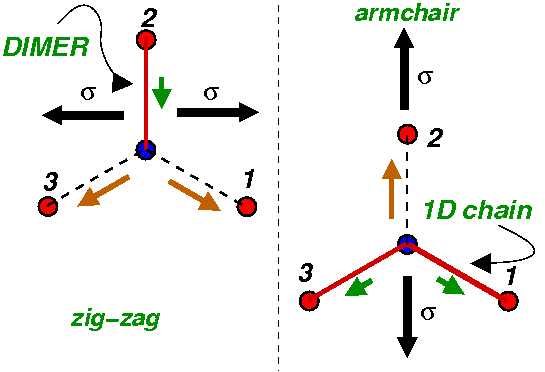}
  \caption{(Color online)
    Representation of the effect of stress on the length
    of the nearest neighbors carbon atoms. On the left we depict the
    ${\cal ZZ}$ case, and, on the right, the ${\cal AC}$ one.
    For the ${\cal ZZ}$ case, the stress $\sigma$ induces
    an increase of the hopping associated with the vertical bond,
    due to the Poisson effect. The hoppings 
    associated with bonds 1 and 3 are reduced, and the system tends
    to dimerize at large deformations.
    For tension along ${\cal AC}$, $\sigma$ is oriented along bond 2. In
    this case all hoppings are reduced, but the one associated with bond
    2 decreases more than the other two, leading to a set of quasi
    one-dimensional chains. The quantitative change of the hoppings
    upon stress was studied quantitatively using {\it ab-initio} methods
    in Ref. [\onlinecite{RicardoStrain}].
  }
\label{fig:bonds}
\end{figure}

For hexagonal systems the relation between stress, $\sigma_{ij}$, and
strain, $u_{ij}$ ($i,j=x,y,z$), reads \cite{sakata}
\begin{equation}
 \left[
\begin{array}{c} 
u_{xx}\\
u_{yy} \\
u_{zz} \\
u_{yz} \\
u_{zx} \\
u_{xy}  
\end{array}
\right]
\!\!=\!\!
\left[
\begin{array}{cccccc} 
S_{11} & S_{12} & S_{13} & 0 & 0 & 0\\
S_{12} & S_{11} & S_{13} & 0 & 0 & 0\\
S_{13} & S_{13} & S_{33} & 0 & 0 & 0\\
0 & 0 & 0 & S_{44} & 0 & 0\\
0 & 0 & 0 & 0 &S_{44}  & 0\\
0 & 0 & 0 & 0 & 0  & 2(S_{11}\!-\!S_{12})\\
\end{array}
\right]
\!\!\left[
\begin{array}{c} 
\sigma_{xx}\\
\sigma_{yy} \\
\sigma_{zz} \\
\sigma_{yz} \\
\sigma_{zx} \\
\sigma_{xy}  
\end{array}
\right]
\end{equation}
where the elements $S_{i,j}$ (here $i,j=1,2,3,4$) are termed compliance
constants.
For the case of graphene under uniaxial tensile strain, the
relation between stress and strain is
\begin{eqnarray}
 u_{xx}=S_{11}\sigma_{xx}\,,\\  
u_{yy}=S_{12}\sigma_{xx}\,, 
\end{eqnarray}
meaning that graphene behaves as an isotropic elastic medium.
We shall consider two cases of stress:  applied along the
zig-zag edges (${\cal ZZ}$), and  applied along the
armchair (${\cal AC}$) edges. In these two cases, the
absolute values of the next-nearest-neighbor vectors $\delta_i$
change as\cite{StressPereira}
\begin{eqnarray}
 |\bd_{1,3}| &=& 1 + \frac{3}{4}\ce -\frac{1}{4}\ce\nu\,,\\
|\bd_2| &=& 1 - \ce\nu\,,
\end{eqnarray}
for the ${\cal ZZ}$ case, and
\begin{eqnarray}
|\bd_{1,3}| &=& 1 + \frac{1}{4}\ce -\frac{3}{4}\ce\nu\,,\\
  |\bd_2| &=& 1 + \ce\,,
\end{eqnarray}
for the ${\cal AC}$ case, where $\ce=S_{11}\sigma$ is the
amount of longitudinal strain, and $\nu=-S_{12}/S_{11}$ is the Poisson
ratio. The two cases correspond to two different physical situations, which can
be understood in the case of extreme deformations using a simple picture:
in the ${\cal ZZ}$ case the system tends to dimerize,
since  $|\bd_{1,3}|$ lengthen, and $|\bd_2|$ shortens; in the ${\cal AC}$
case, all three distances lengthen, but $|\bd_2|$ lengthens more, which can be
 construed as a tendency for the formation of quasi one-dimensional
structures. A real space picture representing these two situations can be seen
in Fig. \ref{fig:bonds}.

\begin{figure}[tb]
  \centering
  \includegraphics*[width=0.45\textwidth]{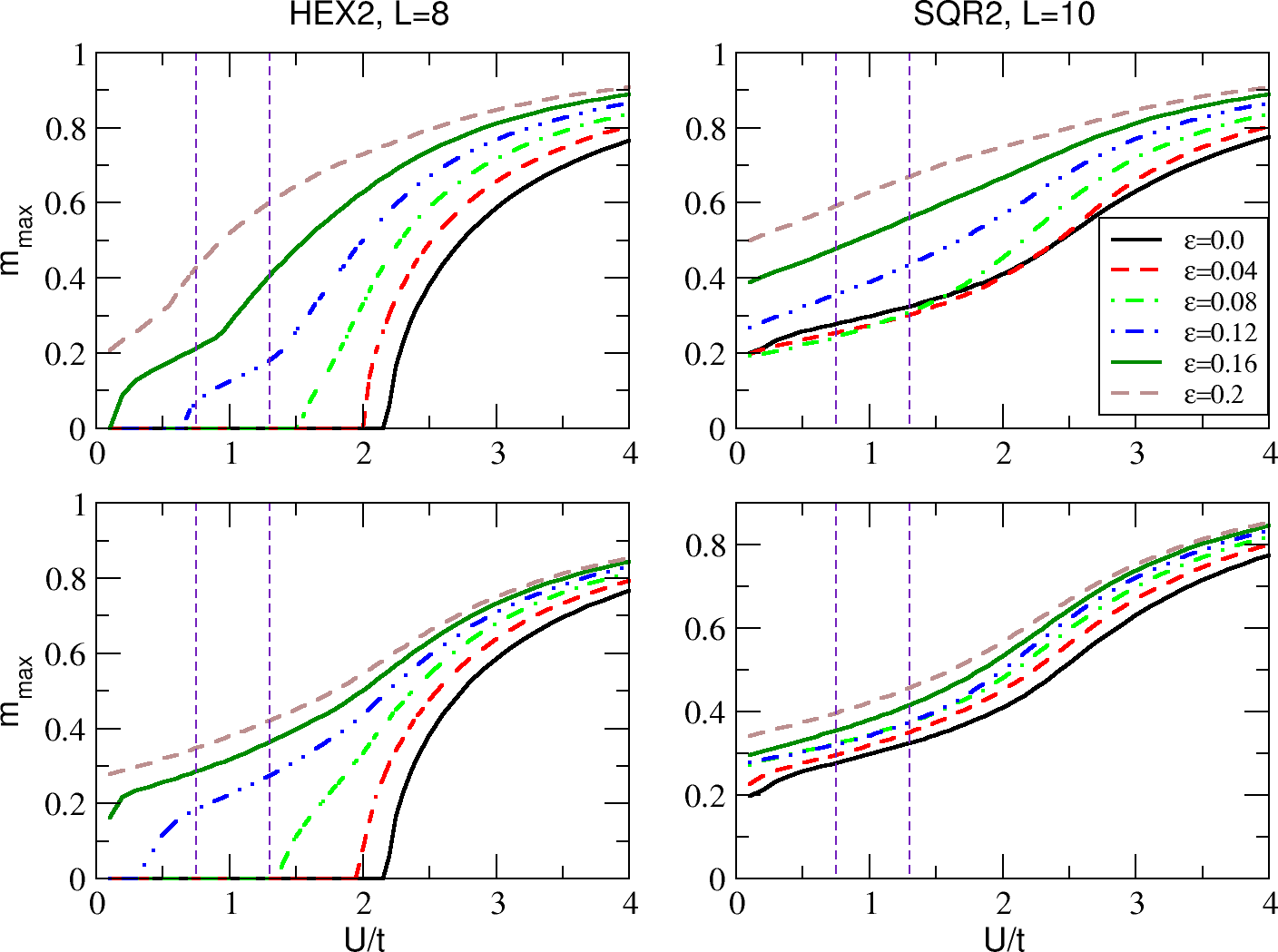}
  \caption{(Color online) 
    Variation of the magnetization as function of the
    Hubbard interaction $U$, for different strain values $\ce$.
    The top panels refer to stress along the ${\cal ZZ}$ edge and the
    other to stress along the  ${\cal AC}$ edge. The Poisson ratio
    used was that of a PET substrate ($\nu\simeq 0.3$), considered in study of
    the Raman red shift of the $G$ and $2D$ peaks of graphene \cite{Lombardo}. 
    The vertical dashed lines refer
    to the values of $U$ used in Refs. \onlinecite{Oleg,Jung} (see text in Sec.
\ref{model} for a discussion about these choices).
  }
\label{fig:strain}
\end{figure}

In Fig. \ref{fig:strain} we present results regarding the effect of strain on
the magnetization of the dots. For the purpose of illustration we consider
dots of smaller size, but the results of Fig. \ref{fig:magU}
guarantee that we should have negligible finite-size effects.
The global effect of tensile stress on the
edge magnetization of the dot is an increase of its magnitude,
independently of whether we consider stress along ${\cal ZZ}$ or
${\cal AC}$ directions. In quantitative terms, the magnetization
increase is more pronounced when stress is applied in the ${\cal ZZ}$
configuration than on the ${\cal AC}$ one. In experimental terms, the prediction
is that magnetism in graphene-based systems should be easier to
detect when the material in under stress.

\begin{figure}[tb]
  \centering
  \includegraphics*[width=0.45\textwidth]{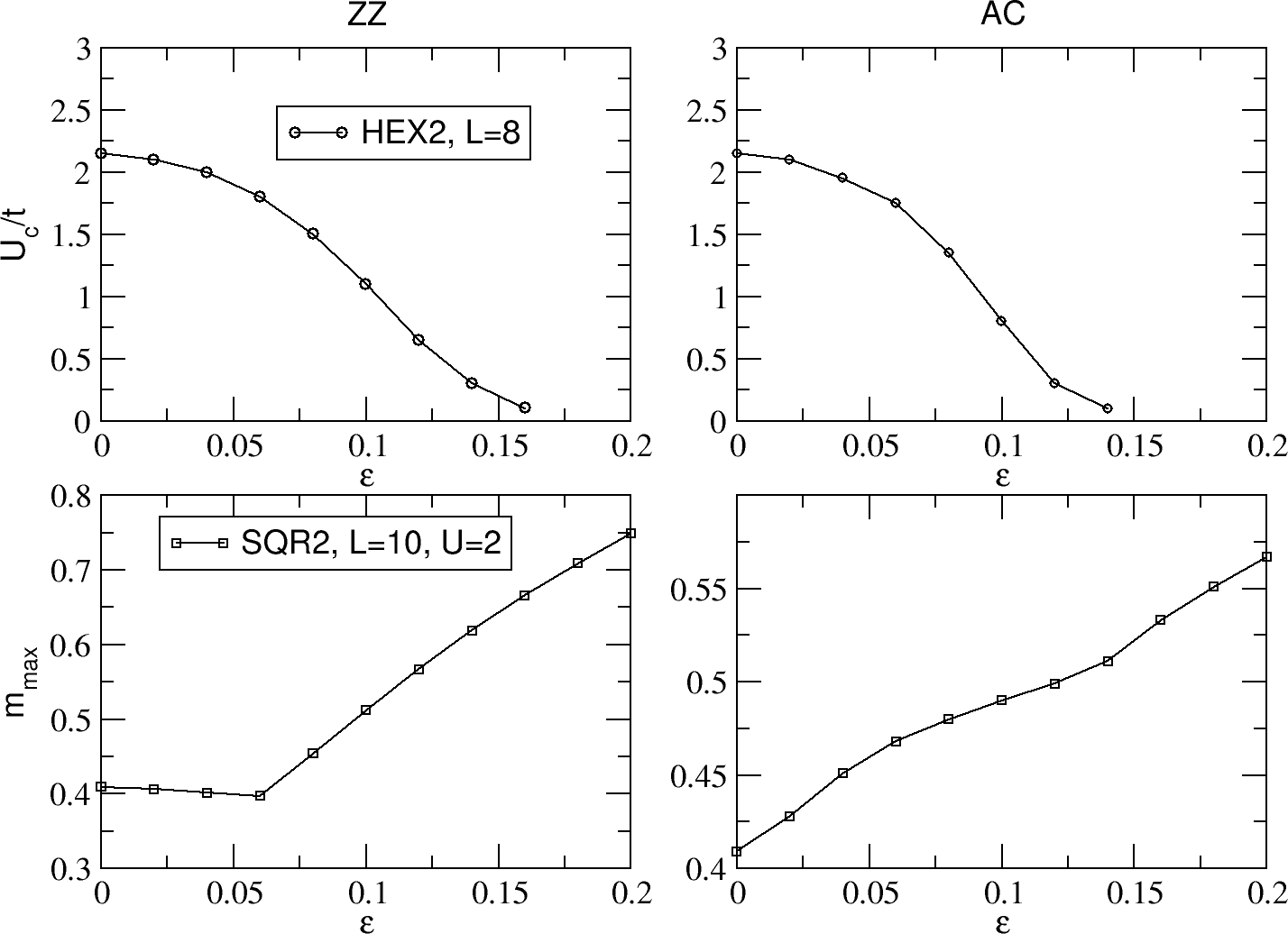}
  \caption{(Color online) 
    Variation of the critical value, $U_c$, with the amount of strain,
    for the cases ${\cal ZZ}$ (top left) and ${\cal AC}$ (top right).
    In both cases we used HEX2-type hexagons with $L=8$. In the bottom
    row we have the dependence of the maximum value of magnetization,
    $m_{\rm max}$, on the amount of strain, using $U/t=2$, and for the
    cases ${\cal ZZ}$ (down left) and ${\cal AC}$ (down right). In
    both cases we used SQR2-type squares, with $L=10$.
  }
  \label{fig:Uc_strain}
\end{figure}

Figure \ref{fig:Uc_strain} shows the explicit variation of $U_c$ with strain,
for the hexagons of type HEX, which have a finite $U_c$. In the same figure, 
on the bottom row, the effect of $\ce$ on the maximum of the magnetization,
$m_{\rm max}$, is also represented for squares of the type SQR2 (which have no
critical $U$). We observe an increase of $m_{\rm max}$
as $\ce$ increases. The dependence of $m_{\rm max}$ on $\ce$ is not the same for
the ${\cal ZZ}$ and ${\cal AC}$ cases at the same $U$ value. Stress along the
zig-zag edge is most effective at producing an enhancement of $m_{\rm
max}$.
This behavior can be understood on the basis of the qualitative physical picture
described in Fig. \ref{fig:bonds}: stress along zig-zag edges tends to produce
dimmers weakly coupled between them, which favors the magnetic state
at those tightly bound atoms.

For hexagons with a finite $U_c$, Fig.~\ref{fig:strain} shows that the
overall effect of strain along both the ${\cal ZZ}$ and ${\cal AC}$ cases is to
reduce the value of $U_c$, which for large $\ce$ obeys $U_c\ll t$. 
At first sight this result may seem easy to understand: 
the value of $U$ cannot change with stress because it is a local (on-site)
property \cite{Endnote-1}.
The hopping, on the other hand, depends strongly on the inter-atomic
distance,
and hence on the external stress. Since the result for $U_c$ in the bulk system,
Eq.~\eqref{Uc}, is bound to the value of the (uniform) hopping $t$, a change in
$t$ produces a change in the absolute value of $U_c$. If we had an uniform
magnetic ground state (as would be the case for bulk graphene without zig-zag
edges), the effect of external stress could be captured through the average
hopping $\tav$, which diminishes as $\ce$ increases, causing a reduction
of the critical value of the Hubbard interaction, $U_c(\ce)\sim \alpha
\langle t(\ce) \rangle$
[$\alpha$ should be around 2.23, as per eq.~\eqref{Uc}]. Since we measure
energies in units of the bare $t$, the above can be written as 
\begin{equation}
  \frac{U_c(\ce)}{t} \sim \alpha \frac{\langle t(\ce) \rangle}{t} 
  \label{eq:Uc-vs-tav}
  \,,
\end{equation}
and thus $U_c(\ce)/t $ would be expected to follow the variation of
$\tav$ with strain.

To verify to which extent such effects contribute to the results shown in
Fig.~\ref{fig:strain} we have calculated the critical Hubbard interaction
expected for a uniform graphene system, as a function of magnitude and
direction of strain. Within the Hartree-Fock framework $U_c$ is given by
\cite{Sorella}
\begin{equation}
\frac{1}{U_c} = \frac{1}{N} \sum_{\bk} \frac{1}{|E(\bk)|} 
\end{equation}
where $N$ is the total number of carbon atoms, and $E(\bk)$ the non-interacting
electron dispersion. In the presence of strain we will have a generalized
dispersion given by
\begin{equation}
  E(\bk) = \pm
  \left|t_2+t_3\, e^{-i\bk.\bm{a}_1}+t_1\, e^{-i\bk.\bm{a}_2}\right|
  \label{eq:Bandstructure}
  \,,
\end{equation}
reflecting that the nearest neighbor hoppings, $t_i$, can all be different in
general \cite{StressPereira}. Using the parameterization introduced in
reference \onlinecite{StressPereira}, we have extracted $U_c$ as a function of
strain magnitude, $\ce$, and orientation with respect to the honeycomb
lattice, $\theta$ ($\theta=0$ for \ZZdir, and $\theta=\pi/2$ for \ACdir). 

\begin{figure}[tb]
  \centering
  \includegraphics[width=0.45\textwidth]{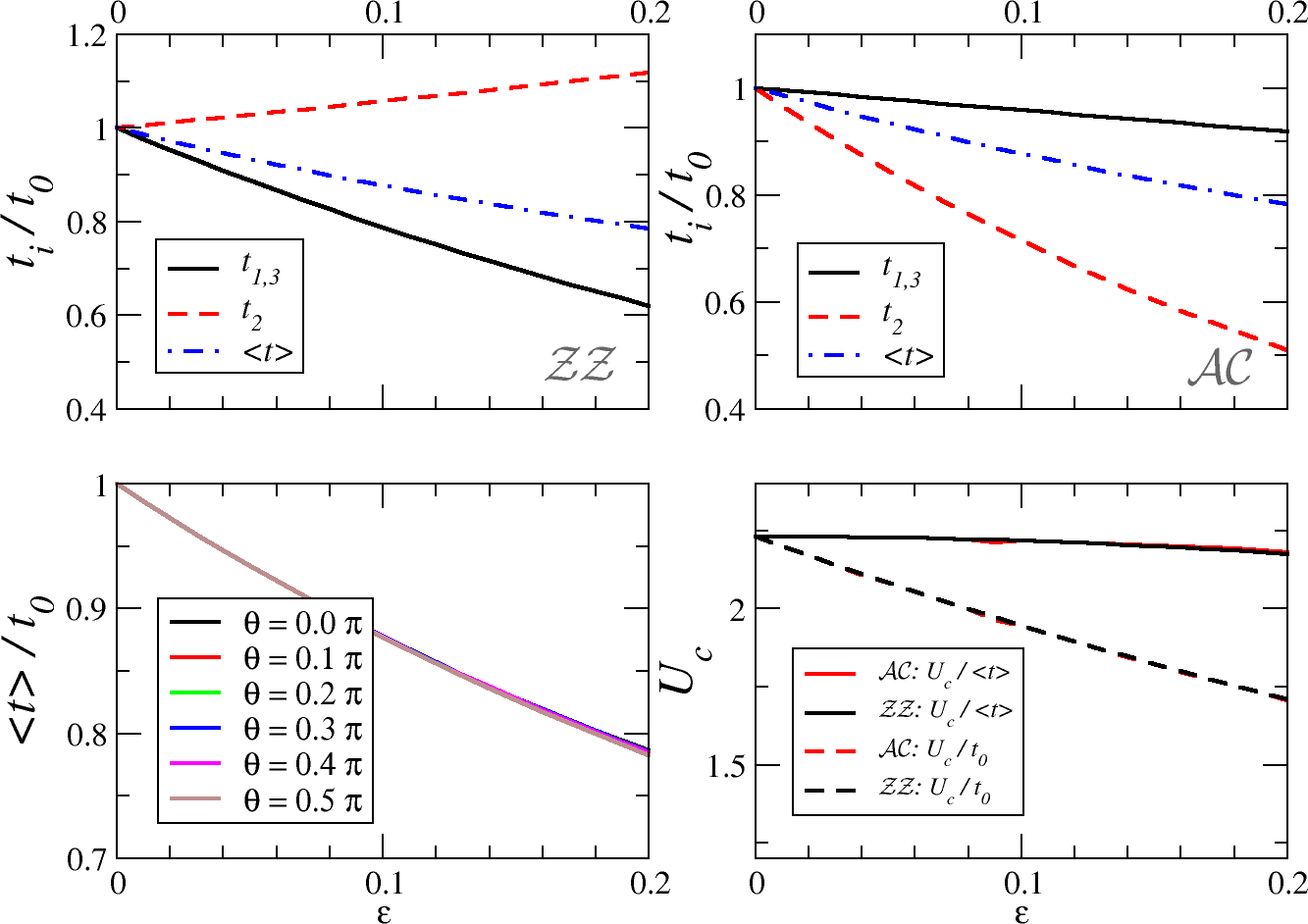}
  \caption{(Color online) 
    The top row shows the variation of the nearest neighbor hoppings
    $t_{1,2,3}$ and the average hopping, $\tav$ under uniaxial strain,
    with strain applied along the \ZZdir\ (left) and \ACdir\
    directions (right). The bottom left panel consists of the
    variation of $\tav$ with strain, for different orientations of the
    uniaxial deformation. In the last panel on the bottom right we
    present $U_c/t_0$ and $U_c/\tav$ for the two representative
    directions.
  }
  \label{fig:hopping}
\end{figure}

Figures \ref{fig:hopping}(a,b) show how the three hopping integrals
$t_1$, $t_2$ and $t_3$ vary under uniaxial strain along the \ZZdir\
and \ACdir\ directions, respectively. Also included in those panels,
are the respective average values of the hopping, defined as the
arithmetic mean of the three nearest neighbor hopping integrals. It
can be seen that the average hopping, $\tav$, is essentially the same
for both \ZZdir\ and \ACdir. This is shown more clearly in
Fig.~\ref{fig:hopping}(c), where we plot $\tav$ as a function of
strain, and for different strain directions: there is no sensible
modification of $\tav$ as the angle $\theta$ defining the tension
direction is changed. Notwithstanding, the tendency is for $\tav$ to
decrease, as we naturally expect. The critical values of $U_c$ under
strain are shown in Fig.~\ref{fig:hopping}(d), where we plot both $U_c
/ t_0$ (that reflects the absolute variation in the critical
coupling), and $U_c/\tav$ (which reflects the statement in
eq.~\eqref{eq:Uc-vs-tav}).
On the one hand, the fact that $U_c/\tav$ is roughly constant up to
deformations of 20\% tallies with the assumption in
eq.~\eqref{eq:Uc-vs-tav} using a constant parameter $\alpha$. However,
even though the decrease in $U_c/t_0$ with strain shown in
Fig.~\ref{fig:hopping}(d) is qualitatively in agreement with the
discussion above regarding the behavior of $U_c$ for the dots in
Fig.~\ref{fig:Uc_strain}, the curves in Fig.~\ref{fig:hopping}(d) do
not decrease as rapidly. Hence, the above argument that the critical
$U$ should follow the variation of $\tav$ \eqref{eq:Uc-vs-tav}, is not
quantitatively accurate. 

\begin{figure}[tb]
  \centering
  \includegraphics*[width=0.45\textwidth]{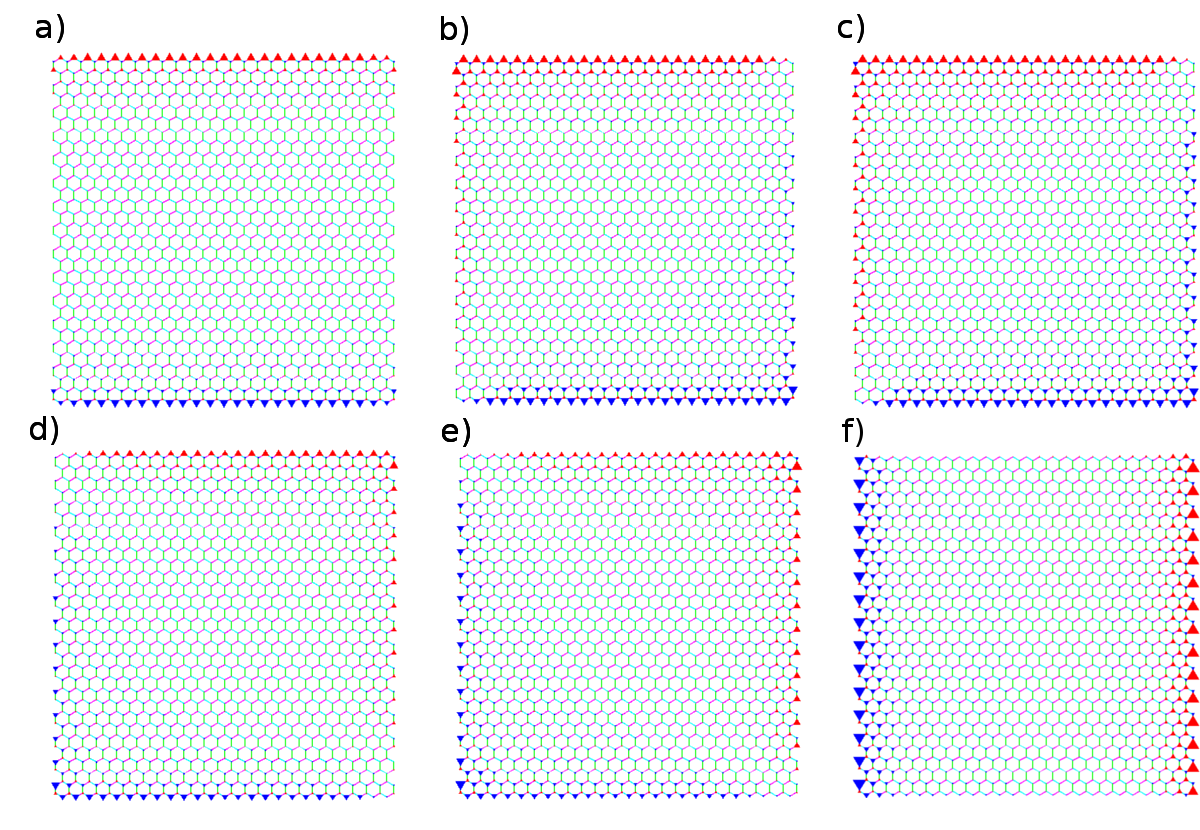}
  \caption{(Color online) 
    Spatial variation of the magnetization as function of 
    strain. (a) $\ce=0$; (b) and (c) corresponds to $\ce=0.14,0.22$, with
    stress along the armchair edge; (d), (e), (f)
    corresponds to $\ce=0.10,0.14,0.22$, with
    stress along the zig-zag edge. In both cases there is an increase
    of the magnetization along the armchair edge as $\ce$ increases.
    The upright triangles represent positive magnetization and the
    down ones represent negative values.
  }
  \label{fig:strain:real}
\end{figure}

The reason for this lies in the very nature of the magnetic ground
states of the quantum dots, which are not uniform. Consequently, the
above argument fails in quantitative accuracy, because it assumes
uniformity. Similarly to what happens in nanoribbons, the
magnetization in the dots studied here has a strong space dependence,
being highly enhanced near certain edges. This is a direct consequence
of the character of the electron states around the Fermi energy, which
tend to be localized near the boundaries. Moreover, as the dots are
deformed by the applied strain, this space distribution is affected as
well. In Fig. \ref{fig:strain:real} we show the spatial evolution of
the magnetization at the edges of the dots of type SQR2 as $\ce$
increases. We have chosen representative values of $\ce$ such that
the effect is clearly evident. Stress along either the armchair edge
(panels (b) and (c) in Fig. \ref{fig:strain:real}), or the zig-zag
edge (panels (d), (e) and (f) in Fig. \ref{fig:strain:real}) shows the
same trend: a tendency for magnetization transfer from the zig-zag to
the armchair edge of the dot. The effect is more pronounced for
tension along the \ZZdir\, as shown in
Figs.~\ref{fig:strain:real}(d-f). In all cases, the value of the
magnetization increases, as can be seen in the lower panels of Fig.
\ref{fig:Uc_strain}. 
From the picture described in Fig. \ref{fig:bonds}, the behavior of
the magnetization in the $\cal{ZZ}$ case can be understood as follows:
large stress along zig-zag edges tends to produce quasi-dimmers,
weakly connected to each other; the dimers at the armchair edges are
only coupled to the bulk of the dot by two weak bonds, and this favors
the stabilization of a magnetic state. 
%

\section{Discussion}
\label{sec:disc}

In this paper we have studied magnetism in graphene quantum dots of
particular geometries, with and without zig-zag edges. Similarly to
what has been proposed in the context of graphene nanoribbons, the
magnetism displayed by these systems may be used in spin filters. We
have not studied the effect of leads on the ferromagnetic order, but
the broadening of the quantum dot levels due to coupling to the leads
(a system with a continuous spectrum), will have an impact on the
magnetic properties of the dot. In the case of a ferromagnetic
nanoparticle, a theoretical description has already been developed
\cite{basko}, but no such model exists for graphene to our best
knowledge. This study will be the subject of a forthcoming
publication. Another aspect we have not considered in our study is the
effect of the substrate on the magnetic properties of the dot. Since
graphene dots are meant to be used as nanoelectronic devices, they
will always interact with some substrate, which can reduce thermal
fluctuations, and favor magnetism at finite temperatures.

It is worth stressing again the main reason for magnetism at the edges
of some graphene systems. Thinking about graphene ribbons or dots as
a bulk system, that is by looking at the total density of states,
leads immediately to the objection that magnetism should not be
present for any finite $U$ value. However, by looking at our results
for magnetism in these systems it is clear that this is a property
of the edges. Therefore the relevant quantity is not
the bulk density of states, but rather the local density of states,
and this latter quantity, near the edges, does become very large at
the Fermi energy ($E=0$), thus leading to very small
critical $U$ values (eventually indistinguishable from zero). This
shows that the total density of states is not a relevant quantity for
this problem.

{\it Ab-initio} calculations have shown that very small benzene-based
systems, such as bisanthrene, have ferromagnetic ground states, and
therefore we expect that magnetism will also be present in small
quantum dots of graphene, as hinted by our Hartree-Fock results. We
have also seen how stress might affect the magnetic ground states.
When applied along the zig-zag edges, stress seems to promote a
spatial rearrangement the magnetization distribution throughout the
dot. This, combined with the transport response of these systems, may
allow mechanical control over spin-polarized currents flowing inside
the dot. On the other hand, the fact that stress along zig-zag edges
leads to the formation of dimers weakly coupled between them, suggests
that the system may prefer to form a sort of spin liquid.
Investigations of whether the ground state will be truly magnetic or
a spin liquid are required. 

In our calculations including strain, we have resorted to linear
elasticity to describe the lattice deformations, as reported in
reference \onlinecite{StressPereira}. We now wish to justify its 
validity, since it is an important point an deserves a careful
explanation. We do not expect linear theory to remain valid for
deformations as high as the 20\% used in some of our calculations
above. However, since we are interested solely in the influence of
strain in the electronic structure, and not on the detailed elastic
response, the relevant detail is not the validity of the linear
theory itself, but rather how a certain amount of strain changes the
hoping values. In our calculations we combined linear theory with the
widespread parameterization of the dependence of the $V_{pp\pi}$ on
the carbon-carbon distance
\begin{equation}
  V_{pp\pi}(l) = t e^{-3.37(l-1)}\,,
  \label{eq:Vpppi}
\end{equation}
where $l$ is the length of the stretched (or compressed) carbon-carbon
distance, in units of that undeformed distance. From the above
equation what matters really is the quantity $(l-1)$, which is
determined by the amount of strain, and its prefactor in the
argument of the exponential. Since there might be legitimate
doubts with respect to the use of linear elasticity for
determining $l$ in the above equation, this issue was
addressed in another publication \cite{RicardoStrain}. There, first
principles calculations were used to study the effect of strain on
the nearest-neighbor hopping values. By their nature, first principles
calculations make no use of any elastic approximation, being valid for
arbitrary deformations, in principle. It was found that the combined
use of linear theory and of the above formula for $V_{pp\pi}$
gives accurate results for the hoping values. Moreover, as discussed
in reference \onlinecite{StressPereira} subsequent \emph{ab-initio}
calculations have shown good quantitative agreement with the use of
linear elasticity combined with the parameterization \eqref{eq:Vpppi}
(for example the merging of the two Dirac points for tension along
the zigzag direction is predicted to occur at the same values of
strain, around 25\%, both \emph{ab-initio} and within tight-binding
with linear elasticity). Therefore our approach to the calculation of
the hopping variation upon strain, and its consequences for the
$\pi$-bandstructure is accurately captured by the linear theory.

The impact of edge roughness on the magnetic properties of the dots
also deserves some considerations. In graphene nanostructures prepared
with current fabrication techniques, the edges are invariably rough
and disordered. This can hinder the stability of long range magnetic
order. However much is still unknown with respect to the nature of
edge reconstruction in real graphene nanostructures. It is expected
that the chemical bonds at the edges be under (surface) tension, which
leads to edge reconstruction as a means of relieving elastic energy.
This reconstruction could lead to self-organization of the edge,
reducing the roughness and allowing for long range magnetic order.
Moreover, the recent advances in tailoring, at the atomic level,
sub-nanostructures by transmission electron microscope
\cite{1DchainsIijima}, might allow unprecedented control over the
edge profiles. Having such control will certainly add greater
tangibility to the prospect of tailoring magnetic states in graphene
nanostructures.

Finally, we point out that the experimental observation of
ferromagnetism arising from grain boundaries with zig-zag edges in
graphite strongly suggests that dimensionality is not a paramount
issue in carbon-based systems, as long as some mechanism for
quenching the thermal and quantum fluctuations is present. Besides
the anisotropy in the interaction between the
local magnetic moments \cite{Magnetism-naturephys}, the fact that
graphene is generally deposited onto a substrate could provide an
extra quenching mechanism, in the same way that it suppresses the
fluctuation-induced crumpling of the two-dimensional graphene
membrane. The mechanical stability of graphene and the robustness of
its magnetic phases can be seen as a vivid example of fluctuation
quenching.

\section*{Acknowledgments}

We thank Antonio H. Castro Neto for stimulating discussions.
N.M.R.P. was supported by FCT under Grant No. PTDC/FIS/64404/2006.
V.M.P was partially supported by the U.S. DOE under the grant
DE-FG02-08ER46512.



\begin{thebibliography}{99}
\bibitem{Endnote-1}
We have discussed in Sec.~\ref{model} how the Hubbard $U$ should be
understood as an effective interaction term. In this sense, strain can
in principle modify the non-local contributions. However, the local
terms are always dominant, and so, as a first approximation, we can
say that deformations of the lattice do not affect sensibly the value
of $U$.
\end{thebibliography}
\end{document}